\documentclass[prl,twocolumn,showpacs,superscriptaddress,preprintnumbers]{revtex4-1}

\usepackage{amssymb,amsmath}
\usepackage{graphicx}

\newcommand{\be}{\begin{equation}}
\newcommand{\ee}{\end{equation}}
\newcommand{\ba}{\begin{eqnarray}}
\newcommand{\ea}{\end{eqnarray}}
\newcommand{\as}{\alpha_{\mathrm{s}}}
\newcommand{\Qs}{Q_{\mathrm{s}}}
\newcommand{\gev}{\;\mathrm{GeV}}
\newcommand{\tev}{\;\mathrm{TeV}}

\newcommand{\lqcd}{\Lambda_{\rm QCD}}
\newcommand{\luv}{\Lambda_\mathrm{UV}}
\newcommand{\fm}{\;\mathrm{fm}}

\newcommand{\bo}[1]{\boldsymbol{#1}}
\newcommand{\xpar}{x_{\parallel}}
\newcommand{\ppar}{p_{\parallel}}

\newcommand{\boXperp}{\bo{x}_\perp^{~}}
\newcommand{\boXperpPrime}{\bo{x}^{\,\prime}_{\perp}}
\newcommand{\Nc}{N_\mathrm{c}}
\newcommand{\cF}{C_\mathrm{F}}
\newcommand{\mT}{m_\mathrm{T}}

\newcommand{\Tr}{\mathrm{Tr}}
\newcommand{\etas}{\eta_\mathrm{s}}
\newcommand{\ybeam}{y_\mathrm{beam}}
\newcommand{\etaflat}{\eta_\mathrm{flat}}

\begin{document}

\preprint{INT-PUB-13-041}

\title{Rapidity Profile of the Initial Energy Density in Heavy-Ion Collisions}

\author{\c{S}ener \"{O}z\"{o}nder}
\email{ozonder@uw.edu}
\affiliation{Institute for Nuclear Theory, University of Washington, Seattle, WA 98195, USA}
\affiliation{School of Physics and Astronomy, University of Minnesota, Minneapolis,
MN 55455, USA}

\author{Rainer J. Fries}
\email{rjfries@comp.tamu.edu}
\affiliation{Cyclotron Institute and Department of Physics and
Astronomy, Texas A\&M University, College Station, TX 77843, USA}


%%%%
%%%%
%%%%
\begin{abstract}
The rapidity dependence of the initial energy density in 
heavy-ion collisions is calculated from a three-dimensional 
McLerran-Venugopalan model (3dMVn) introduced by Lam and Mahlon.
This model is infrared safe since global color
neutrality is enforced.
In this framework,
the 
nuclei have non-zero thickness in the longitudinal direction.
This leads to Bjorken-$x$ dependent unintegrated gluon distribution
functions, which in turn result in a 
rapidity-dependent initial energy density after
the collision. 
These unintegrated distribution functions are substituted in the initial energy density expression
which has been derived for the boost-invariant case. 
We argue that using three-dimensional ($x$-dependent) unintegrated 
distribution functions together with the boost-invariant energy 
formula is consistent given that the overlap of the two nuclei lasts less than the 
natural time scale for the evolution of the fields (${1/\Qs}$) after the collision.
The initial energy density and its rapidity dependence are important 
initial conditions for the quark gluon plasma and its hydrodynamic evolution.
\end{abstract}

\maketitle

%%%%%
%%%%%
%%%%%
\section{Introduction}

In high energy heavy-ion collisions at the Relativistic Heavy Ion Collider
(RHIC) and the Large Hadron Collider (LHC) 
a strongly interacting quark gluon plasma (QGP) has been observed 
\cite{Adcox:2004mh,Adams:2005dq,Muller:2012zq}. 
The initial state of these collisions can be pictured as very strong
classical color fields stretched between the colliding nuclei.
In the color glass condensate (CGC) framework, 
the initial classical color fields can be calculated from the wave function
of the nuclei \cite{McLerran:1993ni,McLerran:1993ka,Kovchegov:1996ty,Iancu:2003xm}. The nuclear wave function at high energies is dominated by
gluons with small momentum fraction $x$, which are radiated from the
partons at large $x$.
The high occupation numbers that the small-$x$ gluons reach at
high energies allow us to use the classical 
color
fields as an
approximation to quantum chromodynamics. 

The classical glue field 
$A^a_\mu$ is 
the
solution of the 
Yang-Mills equation
with the large-$x$ partons of an 
ultrarelativistic nucleus being the source.
From $A^a_\mu$ one can calculate the unintegrated gluon
distribution (UGD) of the nucleus.
This quantity is the main ingredient for the calculation of the initial 
energy density of the QGP and gluon production from classical color fields.
The collision of two nuclei proceeds
through the interaction of their glue fields.
In the light-cone limit the (color-)electric and magnetic fields
$\mathbf{E}^a$ and $\mathbf{B}^a$ in the nuclei,
which
follow from the gauge potential $A^a_\mu$, are transverse.
The longitudinal electric and magnetic fields are then 
initially formed between the nuclei during the interaction
while the transverse field modes between the nuclei are initially zero 
and then grow linearly with time
\cite{Kovner:1995ja,Kovner:1995ts,Fries:2006pv,Lappi:2006fp,Chen:2013ksa}.
The energy that will eventually be available as thermal energy and 
collective kinetic energy of the QGP is deposited in these initial 
longitudinal fields.
Quarks and gluons are produced by the decay of these color 
fields, and the subsequent local thermalization 
of quarks and gluons leads to the formation of QGP.

The realization that the QGP after its formation 
behaves almost like an ideal fluid has made relativistic
hydrodynamics a useful model to describe its expansion and cooling.
Hydrodynamic simulations have proved to be powerful tools in 
analyzing experimental observations \cite{Song:2008hj,Fries:2010ht,Gale:2013da}.
The initial energy density is a key initial condition
that must be specified for these simulations.

Event-by-event and averaged energy densities 
have been calculated before in different
implementations of the CGC approach 
\cite{Fries:2006pv,Lappi:2006hq,Schenke:2012wb,Schenke:2012hg}, 
but mostly in a boost-invariant setup, 
which is valid only around mid-rapidity.
In this approximation,
the colliding nuclei have
been taken to be two-dimensional, infinitely thin sheets rather than 
being longitudinally extended wave packets. The consequences of 
this approximation are $x$-independent UGDs and rapidity-independent 
energy densities. 
In this paper, we calculate the initial energy density $\varepsilon$ at
longitudinal proper time ${\tau\sim 0}$ as a function of rapidity 
by adopting a more realistic picture where the nuclei are slightly
off the light-cone and accordingly have non-zero thickness in the longitudinal 
direction. In the three-dimensional case, a nucleus has partons with
a non-trivial distribution of momentum fractions $x$. 
Collisions of nuclei with $x$-dependent parton distributions lead to 
rapidity-dependent final states.
Here $\varepsilon$ will be ensemble-averaged over all possible 
configurations of the color charge densities of the two nuclei.
The color charge densities fluctuate on an event-by-event basis and
the averaging here corresponds to averaging over multiple events.

In this work,
we employ the  
three-dimensional McLerran-Venugopalan model (3dMVn) 
first developed by Lam and Mahlon \cite{Lam:1999wu,Lam:2000nz}.
Besides the fully three-dimensional treatment of the nuclei,
the 3dMVn model comes with color neutrality enforced on the scale
of a nucleon size, which makes the model well-behaved in the infrared.
The 3dMVn model has two free parameters, 
the strong coupling constant $\as$ and the length scale of
the color neutrality ${\lambda \sim 1.8~{\rm fm}}$. The parameter
space of the model has been explored in Ref. \cite{ozonder,*ozonderErrat} 
through a comparison
between the gluon distribution function calculated from the model
and the parametrization of parton distribution functions from the data by 
Jimenez-Delgado-Reya (JR09) \cite{JimenezDelgado:2009tv}.
Note that our approach is somewhat different from the attempts to calculate 
the rapidity dependence of the initial energy density from UGDs of which the $x$ dependence
is extracted from phenomenological fits or from quantum evolution (see, e.g.,\
Ref. \cite{Hirano:2004en}).

The main result of this work is an expression for the dependence
of the initial energy density on the momentum rapidity $y$. 
We will then make an ansatz to translate the rapidity profile into a 
dependence on space-time rapidity $\etas$.
The initial energy density as a function of $\etas$ is the main input
of these hydrodynamic simulations.

%%%%
%%%%
%%%%
\section{3dMVn Model}

The gluon density in a nucleus at ultra-relativistic
beam energy becomes large, and it is expected to 
saturate at some energy scale due to
gluon recombination.
The saturation density
sets a new dimensional scale ${\Qs \gg \lqcd}$. At the scale $\Qs$
the coupling $\as$ is expected to be weak. However, this does not mean that 
the interactions are weak; on the contrary, the classical fields are strong due 
to the coherence of many small-$x$ gluons.
Therefore, weak coupling techniques can be used to 
understand the structure of the nucleus in spite of its intrinsically non-perturbative
nature.

The equation of motion of Yang-Mills theory is given by
\be
[D_{\mu},F^{\mu\nu}]\equiv\partial_{\mu}F^{\mu\nu}-ig[A_{\mu},F^{\mu\nu}]=J^\nu,
\label{eom}
\ee
where the field strength 
is defined as
\be
F^{\mu\nu}\equiv\partial^{\mu}A^{\nu}-\partial^{\nu}A^{\mu}-ig[A^{\mu},A^{\nu}],
\label{field-strength}
\ee
and $J^\mu$ is a suitably chosen SU(3) current representing
the large-$x$ partons.
As two nuclei, represented by two
currents, pass through each other, their fields can interact through 
the non-Abelian term in the definition of $F^{\mu\nu}$. The result is 
the 
generation of 
chromo-electric ($\mathbf{E}^a$) and chromo-magnetic ($\mathbf{B}^a$) fields 
stretching between the two nuclei. 
In the ultra-relativistic case the longitudinal components dominate 
initially, which is what is used in the original McLerran-Venugopalan (MV)
model \cite{McLerran:1993ni,Lappi:2006fp}. 
In the three-dimensional case, which allows us to calculate 
corrections to this limit, transverse fields are produced initially 
as well, but they are still much smaller than the longitudinal components 
due to Lorentz contraction.
As the nuclei pass through each other at high energies,
some fraction of their kinetic energy is 
deposited in the classical color fields between them. 
This initial energy density as a function of the transverse spatial coordinates as well as rapidity affects the multiplicity, transverse momentum and rapidity distribution of the final particles that reach the detector.

The initial energy density can be calculated from 
the fields via ${\mathcal{H}=\Tr(\mathbf{E}^{2}+\mathbf{B}^{2})}$,
where ${(\mathbf{E},\mathbf{B})=(\mathbf{E}^a,\mathbf{B}^a)t^a}$.
These fields can be written in terms
of the vector potential $A^a_\mu$, which in turn is created by the color charge
densities $\rho_{1,2}^a(\bo{x})$ of the nuclei that enter the current in
Eq.~(\ref{eom}). 
Here we shall use source-averaged quantities in the spirit of 
the original MV
model \cite{McLerran:1993ni}.
A Gaussian measure has been assumed for the ensemble average. The average 
color charge density and its fluctuations at any point in a nucleus are given by 
\cite{Lam:1999wu,Lam:2000nz,ozonder,ozonderErrat}
\be
\langle \rho^a(\bo{x}) \rangle = 0, \label{rho}
\ee
\be
\langle\rho^{a}(0)\rho^{b}(\bo{x})\rangle=\delta^{ab}
\kappa_{A}^{3}  
\left[\delta^{3}(\bo{x})-
\frac{3
\, \exp \! \left( - 
\frac{\sqrt{3} \vert \bo{x} \vert }{ \lambda} 
\right)
}{4\pi\lambda^2
\vert \bo{x} \vert
}
\right],
\label{rhorho}
\ee
where the average
squared color charge per unit volume
is determined by $\kappa^3_A=3A\cF/(\Nc^2-1)V=3A/2\Nc V$ 
(we assume that the valence quarks of the nucleons are the source of all gluons).
The correlation length of valence quarks is set by $\lambda \sim \lqcd^{-1}$. The term in Eq.~(\ref{rhorho}) that includes $\lambda$ mimics confinement through colored noise and hence cures the infrared divergence problem. 
The correlation length is found to be $\lambda \sim 1.8\fm$ from the
comparison between the 3dMVn model and measured gluon distribution functions \cite{ozonder,ozonderErrat}.
In the absence of the $\lambda$-dependent term, the spectrum of the
correlation function in Eq.~(\ref{rhorho}) would resemble that of white
noise. In that case, fluctuations at all scales, including $\vert \bo{q}
\vert<\lqcd$, would exist and this would cause a divergence in the infrared \cite{Lam:1999wu,Lam:2000nz,ozonder,ozonderErrat}. 

The three-dimensional coordinate system ${\bo{x}=(\xpar,\boXperp)}$ to 
be used in the 3dMV model is defined in the rest frame of the nucleus.
The longitudinal coordinate is given by \cite{Lam:2000nz}
\be
\xpar = \frac{1}{\epsilon}x^{-}-\frac{\epsilon}{2}x^{+},
\label{xpar}
\ee
where $\epsilon=\left[2(1-\beta)/(1+\beta)\right]^{1/2}$ and $\beta$ is the 
speed of the nucleus. 
The light-cone coordinates are defined as ${x^{\pm}=(t\pm z)/\sqrt{2}}$.
The longitudinal coordinate $\xpar$ is conjugate to $\ppar$, which is related to
the parton momentum fraction via
\be
\ppar = mx,
\label{ppar}
\ee
where $m$ is the nucleon mass. 

The ensemble-averaged initial energy density ${\varepsilon(\tau=0)}$ can be
written in terms of the correlation functions $\langle
A_{i}^{a}(\bo{x})A_{i}^{b}(\bo{x}^{\,\prime})\rangle$ 
for each nucleus in light-cone gauge.
For a given nucleus, the vector field correlation function in momentum space is given by \cite{Lam:2000nz,ozonder,ozonderErrat}
\ba
\langle A_{i}^{a}(\bo{q})A_{i}^{a}(-\bo{q})\rangle
& = &
12 \pi \as \frac{\Nc^2-1}{\Nc} \frac{ A}{m^2 x^2}
\int d^{2} {\boldsymbol{\Delta}_\perp}
e^{i {\boldsymbol{q}_\perp}\cdot
{\boldsymbol{\Delta}_\perp}} \nonumber \\
& &  \quad
\, \, \times {\cal L}(x;{\boldsymbol{\Delta}_\perp})
{\cal E}\boldsymbol{(}v^{2}L({\boldsymbol{\Delta}_\perp})\boldsymbol{)},
\label{gluon-distro-LM}
\ea
where $A$ is the mass number of the nucleus and
${\bo{\Delta}_\perp=\boXperp-\boXperpPrime}$. Here
${\bo{q}=(q_\parallel,\bo{q}_\perp)}$ is the momentum conjugate to the rest
frame coordinate ${\bo{x}=(x_\parallel,\bo{x}_\perp)}$. The pair distribution
functions $\mathcal{L}$ and $L$ are convolutions of the Green's function 
with Eq.~(\ref{rhorho}), and they are given as
\begin{multline}
\mathcal{L}(x;\bo{\Delta}_\perp)=  - \frac{1}{12 \pi} \Bigg[ (x m \lambda)^2 K_0(x m \Delta_\perp) \\
 \quad -\Big( 3+(x m \lambda)^2 \Big) K_0(\Delta_\perp 
\sqrt{3+(x m \lambda)^2 } / \lambda) \Bigg],
\end{multline}
and
\be
L(\bo{\Delta}_\perp)= - \frac{\lambda^2}{6 \pi} \left[K_0 \left( \frac{\sqrt{3} \Delta_\perp}{\lambda} \right)  + \ln \left(  \frac{\sqrt{3} \Delta_\perp}{2 \lambda} \right) + \gamma_E  \right].
\ee
For a cylindrical nucleus, the nuclear correction factor $\mathcal{E}(z)$ and its argument $v^2$ are given by
\be
\mathcal{E}(z)= \frac{1}{z}(e^z -1),
\ee
and
\be
v^2=\frac{3 A g^4}{2 \pi R_A^2}
\approx 24 \pi \as^2 A^{1/3} 
r_0^{-2}.
\ee

The correlation function in Eq.~(\ref{gluon-distro-LM}) can be related to the Weizs\"{a}cker-Williams unintegrated gluon density (UGD)
\be
\phi(x,\bo{q}^2_\perp) \equiv x\frac{dN}{dxd^{2} \bo{q}_{\perp}} = \frac{m^2 x^2}{4 \pi^3} \langle A_{i}^{a}(\bo{q})A_{i}^{a}(-\bo{q})\rangle.
\label{UGD}
\ee
The UGD in Eq.~(\ref{UGD}) can be expressed in terms of either the longitudinal
momentum $q_{\parallel}$ or momentum fraction $x$ via the relation
${x=q_{\parallel}/ m}$.
The gluon distribution function for the nucleus of mass number $A$ is defined as the integral of the UGD
\be
xg_{A}(x,Q^{2})\equiv \int^{Q^{2}}d^{2} \bo{q}_{\perp}\phi(x,\bo{q}_{\perp}^{2}).
\label{gluon-pdf}
\ee
So far, we have reviewed the three-dimensional
  ($x$-dependent), color neutral 
UGD
  of the 
  3dMVn model. In the next section, we will find the energy density in 
  terms of those distributions.

%%%%
%%%%
%%%%
\section{Rapidity-Dependent Energy Density}

The aim in this section is to express the energy density of the longitudinal
fields, which form after the collision, in terms of the UGDs
$\phi(x,\bo{q}_{\perp}^{2})$ of the nuclei.
In the original MV
model, the interaction of nuclei as
infinitely thin sheets is instantaneous at the hyper-surface 
${\tau=0}$. By imposing the continuity of the vector potential on that hyper-surface via the classical Yang-Mills equation, one finds two boundary conditions \cite{Kovner:1995ja,Kovner:1995ts}.
At $\tau=0$, these boundary conditions
determine the vector potential after the collision
  with light-cone components $(x^+ A, -x^- A, A_{\perp}^{i})$ 
  in terms of the vector potential of the incoming nuclei $A_{1}^{i}$ and $A_{2}^{i}$. Hence, the initial energy density after the collision can be expressed in terms of the vector potential of the incoming nuclei before the collision. 

In the three-dimensional case where nuclei are moving with speed $v<1$,
  the interaction is not instantaneous due to the longitudinal extent of the nuclei. Hence, it is not possible to define a single interaction hypersurface, which makes the interaction time dependent. Also, since the nuclei are not on the lightcone, there will be some longitudinal fields in the nuclei, which would give rise to initial transverse fields right after the interaction. In this work, we
will not fully solve the corresponding problem
  in the 3dMVn setup. Rather, we will rely on the
boost-invariant energy expression (see Eq. (\ref{eq:edens})) which assumes instantaneous interaction at $\tau=0$
and no longitudinal fields in the nuclei before the collision. 
This approximation uses the fact that for sufficiently large energy
the interaction time ${\Delta \tau \sim R_A/\gamma}$, where $\gamma$ is the
Lorentz factor of the nuclei in the laboratory system, is smaller than the 
natural time scale for the evolution of the fields after the collision 
${\sim 1/\Qs}$. Since $d\varepsilon/d\tau = 0$ in the MV model \cite{Krasnitz:1998ns,Chen:2013ksa}, we can argue that $\varepsilon$ does not vary much over time scales $\Delta \tau$
and it will be acceptable to use the formula derived in the MV model for 
$\tau=0$ (see Eq.~(\ref{eq:edens})) as an approximation for the 
energy density just after nuclear overlap.
In addition, the longitudinal fields in the nuclei before the collision 
 in the three-dimensional setup are still suppressed by the large Lorentz boost
and we will neglect the contribution of the transverse fields they generate
on the energy density.
It is important to note, however, that 
  despite these approximations we will use the 3dMVn 
  ($x$-dependent) version of the correlation function of transverse vector 
  fields which are related to the three-dimensional, $x$-dependent UGDs.

Thus from here on, when we refer to the initial energy density we 
refer to the proper time just after the overlap of the two nuclei, 
and we will sometimes quote it as ${\tau \to 0}$.
We use the classic result of the MV model for the initial energy density
from the longitudinal fields after the collision 
\cite{Lappi:2006hq,Makhlin:1996dr,Krasnitz:1998ns}
\begin{equation}
  \varepsilon (\tau\to 0) = \Tr (E^\eta E^\eta+B^\eta B^\eta).
  \label{eq:edens}
\end{equation}
The longitudinal fields after the collision in Eq.~(\ref{eq:edens}) are given in 
terms of the transverse nuclear fields before the collision
in lightcone gauge as \cite{Fujii:2008km,Chen:2013ksa}
\begin{equation}
E^\eta (B^\eta) = ig \delta^{ij} (\epsilon^{ij}) \left[ A_1^i , A_2^j \right],  
\end{equation}
where $E^\eta=E^{a\eta}t^a$. Plugging this into Eq.~(\ref{eq:edens}) leads to
\begin{align}
\varepsilon(\tau \to 0)= \frac{1}{2} g^2 f^{abc} f^{dec} & \, ( \delta^{ij} \delta^{kl} + \epsilon^{ij} \epsilon^{kl})\nonumber \\
& \quad \times \langle A_i^a A_k^d\rangle \langle A_j^b A_l^e \rangle,
\label{energydeltaepsilon}
\end{align}
where the nuclear indices $1$ and $2$ have been dropped for simplicity because
we assume a central collision for which averaged fields will be the same for
both nuclei when they have the same mass number.

The dependence of the initial energy density on the rapidity $y$ 
can be inferred if we use the $x$-dependent 3dMVn UGDs from
Eq.~(\ref{UGD}) in the expression for the energy density. Those UGDs
are summed over color and spatial indices. Equation~(\ref{energydeltaepsilon}), however, includes correlators with most general color and spatial indices. 
For the
latter, we make an ansatz which employs the color neutral, 
$x$-dependent UGD. Our ansatz here is a generalization of the one in Ref. \cite{Lappi:2006hq}:
\be
\langle A_{i}^{a}(\bo{q})A_{j}^{b}(\bo{p})\rangle = \frac{\delta^{ab}}{\Nc^{2}-1} \delta_{\bo{p},-\bo{q}}  
\langle A_{i}^{a}(\bo{q})A_{i}^{a}(-\bo{q})\rangle \frac{p_{\perp i} p_{\perp j}}{\bo{p}_{\perp}^{2}},
\ee
where again ${\bo{q}=(q_\parallel,\bo{q_\perp})}$. Using
$\delta^{3}(\bo{p}-\bo{q})\Leftrightarrow  V \delta_{\bo{p},\bo{q}} /
(2\pi)^{3}$ and substituting 
the diagonal correlation function from Eq.~(\ref{UGD}), we find
\begin{align}
\begin{split}
\langle A_{i}^{a}(\bo{q})A_{j}^{b}(\bo{p})\rangle=& \frac{\delta^{ab}}{\Nc^{2}-1}  \frac{1}{V} (2\pi)^{3}  \delta(p_{\parallel}+q_{\parallel})\delta^{2}(\bo{p}_{\perp}+\bo{q}_{\perp}) \\
& \qquad \times \frac{p_{\perp i} p_{\perp
    j}}{\bo{p}_{\perp}^{2}}\frac{4\pi^{3}}{m^{2}x^{2}}\phi(x,\bo{p}^2_\perp)  .
\end{split}
\label{ansatz-final}
\end{align}
Apart from being three dimensional and $x$ dependent, our ansatz includes a
factor $1/V$ that is missing in the ansatz in Ref. \cite{Lappi:2006hq}. This is
because the UGD in Ref. \cite{Lappi:2006hq} has been defined as per unit area and a
factor of area has been included in the definition of the gluon distribution
function, unlike
the definition
given in Eq.~(\ref{gluon-pdf}).

Writing the fields in Eq.~(\ref{energydeltaepsilon}) in momentum space and using our ansatz given in Eq. (\ref{ansatz-final}) lead to
\begin{align}
\begin{split}
\varepsilon(\tau \to 0) &=
\frac{g^{2}}{2}\frac{\Nc}{\Nc^{2}-1}\frac{1}{V^{2}} \\
& \times \int^{\luv}\frac{d^{3}\bo{p}}{(2\pi)^{3}} \frac{d^{3}\bo{q}}{(2\pi)^{3}} \frac{4\pi^{3}}{p_{\parallel}^{2}}\phi(p_{\parallel},\bo{p}^2_{\perp})\frac{4\pi^{3}}{q_{\parallel}^{2}}\phi(q_{\parallel},\bo{q}^2_{\perp}),
\end{split}
\label{energy-with-LambdaUV}
\end{align}
where $d^3 \bo{p}=d p_\parallel d^2 \bo{p}_\perp$, and $\luv$ is a UV cutoff.
We go from the longitudinal momenta $p_\parallel = x_1 m$ , 
$q_\parallel = x_2 m$ of the partons from both nuclei to the momentum rapidity 
$y$ of the produced gluon via momentum conservation of a $2\to 1$ gluon
fusion process. 
The rapidity of the produced gluons after the interaction and the momentum fraction of the gluons from nucleus $1$ and nucleus $2$ before the collision are related via
\be
x_{1,2}= \frac{ \mT}{\sqrt{s}} e^{\pm y},
\label{x-exp-y}
\ee
where $\mT$ is a scale related to the transverse mass of the 
produced gluon and $\sqrt{s}$ is the center of mass energy per nucleon pair.
The integration over the longitudinal momenta in
Eq.~(\ref{energy-with-LambdaUV}) is rewritten as follows:
\be
d p_\parallel d q_\parallel = m^2 d x_1 d x_2 \longrightarrow d \mT^2 dy \frac{1}{s}.
\ee
Now we can write the
density differential in rapidity as
\begin{align}
\frac{d\varepsilon}{dy}=& \frac{g^{2}}{2}\frac{\Nc}{\Nc^{2}-1}\frac{1}{V^{2}} \left( \frac{4 \pi^3}{2 \pi} \right)^2 \int^{\luv}
\frac{d^{2}\boldsymbol{p}_\perp}{(2\pi)^{2}}\frac{d^{2}\boldsymbol{q}_\perp}{(2\pi)^{2}} \nonumber
 \\
&  \qquad \times
 \frac{1}{m^2} \int d \mT^2
\frac{s}{\mT^4}
\phi(x_{1},\boldsymbol{p}_{\perp}^{2})
\phi(x_{2},\boldsymbol{q}_{\perp}^{2}).
\label{dedy-with-mT}
\end{align}
We have confirmed that the rapidity dependence was not 
affected if the $\mT^2$ integration was replaced by an average 
value of the transverse mass 
$1 \mapsto  \langle \mT^{2}\rangle\delta(\mT^{2}-\langle \mT^{2}\rangle)$ for
simplicity.
Finally, we obtain
\begin{eqnarray}
\frac{d\varepsilon}{dy}&=& 8 \pi^5 \as \frac{\Nc}{\Nc^{2}-1}\frac{1}{V^{2}} 
\int^{\luv}
\frac{d^{2}\boldsymbol{p}_\perp}{(2\pi)^{2}}\frac{d^{2}\boldsymbol{q}_\perp}{(2\pi)^{2}} \nonumber
 \\
 &  & \qquad \qquad \times
 \frac{1}{m^2}  
\frac{s}{\langle \mT \rangle^2}
\phi(x_{1},\boldsymbol{p}_{\perp}^{2})
\phi(x_{2},\boldsymbol{q}_{\perp}^{2}).
\label{dyde-final}
\end{eqnarray}

Starting from Eq.~(\ref{dyde-final}) one can perform a numerical evaluation of
the momentum integrals, or one can derive a pocket formula using the
expressions for the UGD given in Eqs.~(\ref{gluon-distro-LM}) and (\ref{UGD}). After taking 
the integrals and regularizing the UV singularity by coarse graining the limit
$|\boldsymbol{x}_\perp| \to 0$ at a scale $|\boldsymbol{x}_\perp| \sim 
1/\Lambda$, we arrive at 
\begin{multline}
  \frac{d\varepsilon}{dy} =
72 \pi \as^3 A^2 \frac{\Nc^2-1}{\Nc} \frac{s}{m^2 \langle\mT\rangle^2} \frac{1}{V^2}
  \mathcal{E}^2\left( v^2 L(\Lambda^{-1}) \right) \\ \times
  \mathcal{L}\left(\frac{\langle \mT \rangle}{\sqrt{s}}e^{y}; 
  \Lambda^{-1}\right)
  \mathcal{L}\left(\frac{\langle \mT \rangle}{\sqrt{s}}e^{-y}; 
  \Lambda^{-1}\right) ,
  \label{analyticresult}
\end{multline}
where $\Lambda$ is related to $\luv$ through a numerical
constant of order $\mathcal{O}(1)$. One could simplify the analytic result in Eq.~(\ref{analyticresult})
further by expanding the Bessel functions of the second kind in $L$ and 
$\mathcal{L}$, and the exponential function in $\mathcal{E}$ without a
significant increase in the numerical uncertainty for given realistic parameters
for not too large rapidity $y$. 
In the next section, we present our numerical estimates for 
${d\varepsilon / dy}$.

%%%%
%%%%
%%%%
\section{Results}

For our numerical calculations we take ${\Nc=3}$, ${m=1\gev}$ and ${\lqcd=0.2\gev}$. We also assume ${\as=0.4}$ and $\lambda=1.8\fm$ as a result of the 
comparison between the 3dMVn model and the JR09 parametrization of the 
nuclear gluon distribution function carried out in Ref. \cite{ozonder,ozonderErrat}. 
We parametrize the volume of the cylindrical nucleus with mass number $A$ as $V=\pi R_A^2 h$ where the nuclear radius is  
$R_A=r_0 A^{1/3}$ and the longitudinal length (in the rest frame of the
nucleus) is taken to be ${h=R_A}$. For Au and Pb, we take $r_0=1.1\fm$ and 
$r_0=1.3\fm$, respectively. 

\begin{figure}[t]
\begin{raggedleft}
\includegraphics[scale=0.625]{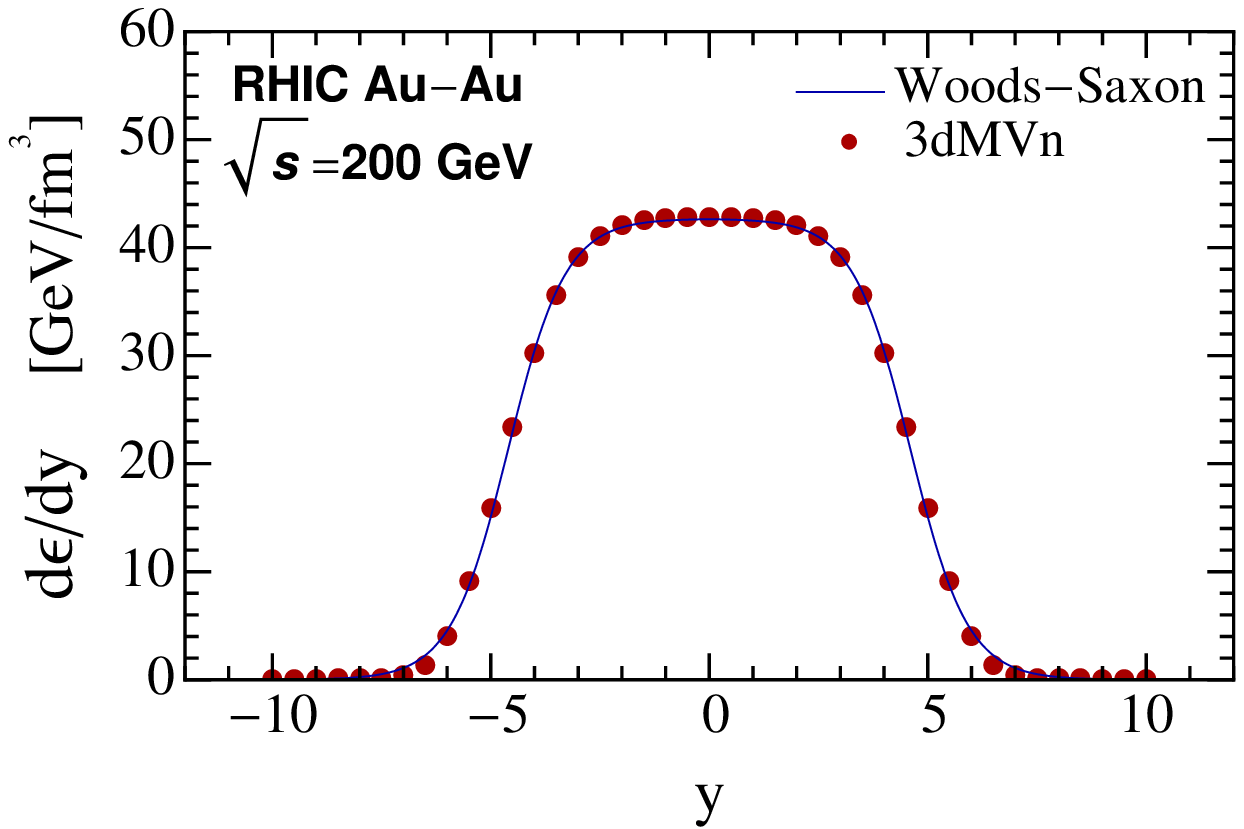}
\par\end{raggedleft}
\vspace{0.5cm}
\begin{raggedleft}
\includegraphics[scale=0.66]{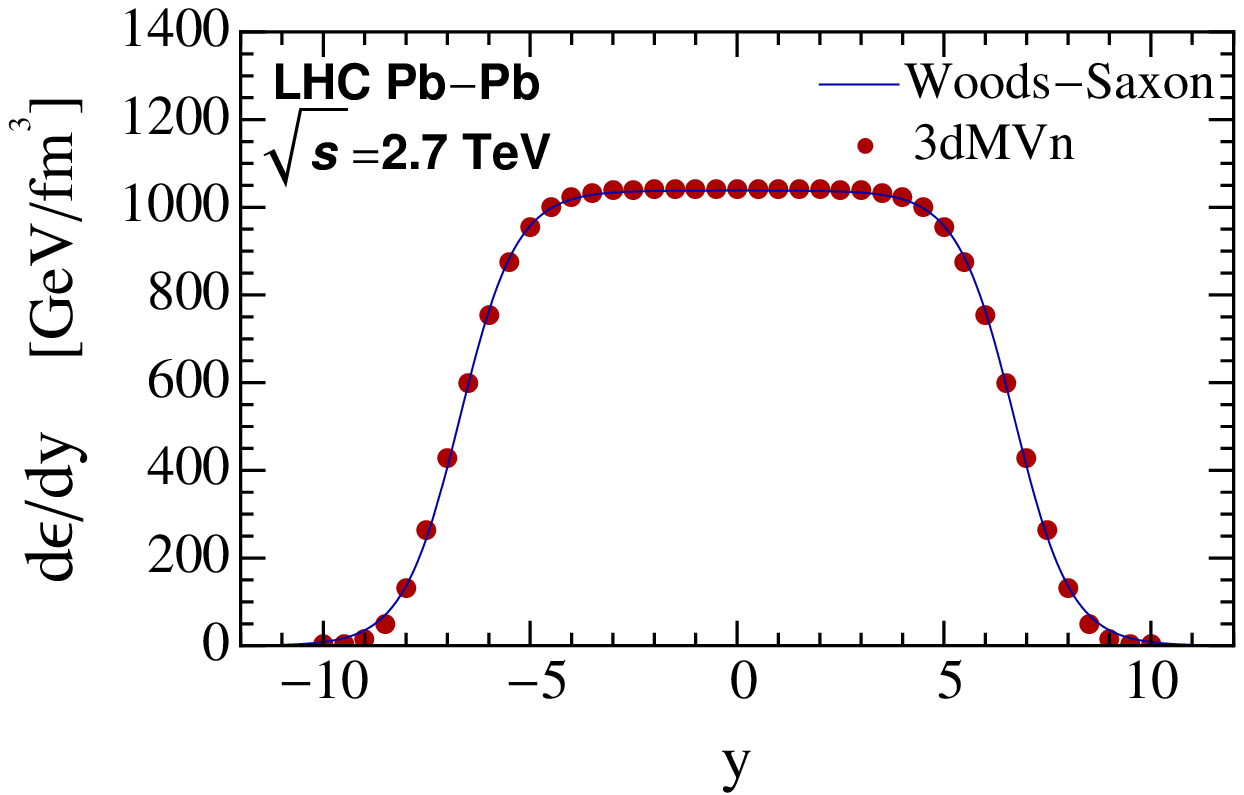}
\par\end{raggedleft}
\caption{(Color online) $d\varepsilon/dy$ as a function of rapidity $y$ for central Au-Au 
 collisions at RHIC (upper panel) and central Pb-Pb collisions at LHC 
(lower panel). 
$d\varepsilon/dy$ can also be taken as $\varepsilon(\etas)$
(see text).
The shape of the energy density from 3dMVn is fitted well 
by a Woods-Saxon profile in both cases. }
\label{RHICandLHC}
\end{figure}

The numerical evaluation of the spatial integration in 
Eq.~(\ref{gluon-distro-LM}) has been carried out 
between the limits ${0\leq \Delta_\perp  \leq 2\fm}$; the
  integrand does not contribute significantly for $\Delta_\perp >
  2\fm$.
For Pb-Pb ($A=207$) at LHC at $\sqrt{s}=2.76\tev$, we use $\luv \sim \Qs \sim 2.5\gev$ 
as the cutoff for
the integral in Eq.~(\ref{energy-with-LambdaUV}). For Au-Au ($A=197$) at RHIC at $\sqrt{s}=200\gev$, we use $\luv \sim \Qs \sim 2\gev$. The transverse momentum of the produced gluons is assumed to be $\langle \mT \rangle=0.7\gev$ and $\langle \mT \rangle=1.3\gev$ for RHIC and LHC, respectively.
Our final results are shown in Fig.~\ref{RHICandLHC}.

Let us recall that the 3dMVn model extends the validity of the MV
model to larger Bjorken-$x$. For Au and Pb nuclei it merges with MV below
$x\sim 0.05$, but it describes UGDs up to $\sim 0.25$ for certain values
of $Q^2$ \cite{Lam:2000nz,ozonder,ozonderErrat}. This means that our
results can be estimated to be reliable to about $y\approx 4$ for 
RHIC energies and $y\approx 6$ for LHC, far beyond the region of validity 
of the original MV model applied to those energies.

Now we turn to the discussion of finding $\varepsilon(\etas)$ from $d\varepsilon/dy$.
Let us recall that 
\begin{equation}
\frac{d\varepsilon}{dy} = \frac{dE}{dy \, \tau d\etas d^2x_\perp},
\label{dEdytaudetadx}
\end{equation}
where $\etas$ is the space-time rapidity. The simple picture of the
transverse structure in the cylindrical 3dMVn model ensures that the 
energy density is homogeneous as a function of $\mathbf{x}_\perp$ in 
the nuclear overlap zone 
(one can certainly go beyond this approximation, see Ref. \cite{Chen:2013ksa}). 
Also, we have not
solved the full non-boost invariant collision problem that 
would give us the $\etas$ dependence. However, we will now postulate a relation between
rapidity $y$ and space-time rapidity $\etas$ that will allow us to estimate
the dependence on $\etas$.
One reasonable ansatz follows 
from the Bjorken flow where we 
consider a Hubble-like velocity profile for the produced gluons, i.e., ${v_z=p_z/E\approx z/t}$ 
\cite{Bjorken:1982qr,Hirano:2004en}.
This leads to ${y \approx \etas}$, where ${\etas \equiv \frac{1}{2} \ln \left[ (t+z)/t-z  \right]}$ and 
${y \equiv \frac{1}{2} \ln \left[ (E+p_z) / E-p_z  \right]}$
Therefore, our ansatz can be summarized as 
\begin{equation}
  \frac{d E}{dy\tau d\etas d^2x_\perp} \approx \varepsilon(\etas) 
  \delta(y-\etas).
  \label{Hubble-ansatz}
\end{equation}
However, the $d\varepsilon/dy$ that we calculated in Eq.~(\ref{dyde-final})
did not have this 
correlation, rather it was flat in $\etas$ for any fixed value of $y$ and
must thus correspond to an averaged quantity
\begin{equation}
  \frac{d\varepsilon}{dy}   \Big|_{y \approx \etas} 
  \propto \int d \etas
  \frac{d E}{dy\tau d\etas d^2x_\perp} \, .
\end{equation}
 This leads to the relation
\begin{equation}
  \varepsilon(\etas) \propto
  \frac{d\varepsilon}{dy}
  \Big|_{y \approx \etas}.
  \label{y-to-eta}
\end{equation}
Thus 
we infer that
the momentum rapidity profile would coincide with the 
space-time rapidity profile. 
However, we will not attempt to find the proportionality constant between
$\varepsilon(\etas)$ and $d\varepsilon/dy$.
This result could be generalized by allowing 
additional smearing around the $y=\etas$ relation.

The normalizations of our results depend only weakly on 
the correlation length $\lambda$, but 
they are sensitive to the choice of $\as$, ${\luv \sim \Qs}$, 
${\langle \mT \rangle}$, and the nuclear volume $V$. However, reasonable values for these parameters lead to 
a normalization
  of $\varepsilon(\etas)$
   which is comparable to
  the typical initial conditions used in hydrodynamics simulations. 
  Our main
  result here is the shape of the rapidity dependence of the initial energy density, which is much more stable against
the variations of the parameters. Note that
the uncertainty grows significantly at large rapidities 
$|\etas| \approx |y| \approx \ybeam$.
Our numerical calculations exhibit a plateau 
  around mid-rapidity and a fall-off toward beam rapidity that is about 
  three
  units of rapidity wide at both energies. The shape is fitted well by a 
  Woods-Saxon profile 
\begin{equation}
  \varepsilon(\etas) = 
  \frac{\varepsilon_0}{1+\exp\left[    (\vert \etas \vert -\etaflat ) / a 
  \right]}
  \, .
  \label{ws}
\end{equation} 
The values of the Woods-Saxon parameters for 
RHIC and LHC 
are given in Table~\ref{table-woods-saxon}.

Our Woods-Saxon parametrization is somewhat different from the usual
phenomenological parametrization of the initial energy density used by most
groups employing (3+1)$D$ hydrodynamics, which consists of a perfectly flat
plateau around mid-rapidity flanked by two half-Gaussian functions 
\cite{Hirano:2001eu,Schenke:2010nt}. 
It remains to be seen whether the differences between our calculated
shape and the empirical assumptions in the literature can be resolved 
experimentally.
However, our calculation makes quantitative predictions of the width of the 
plateau and the width of the fall-off towards the beam rapidity,
albeit within the limitations of the approximations used.

Our results can be compared with the ones in Ref. \cite{Hirano:2004en},
which use different assumptions. In that reference the energy density
is extracted from the number density, which in turn comes from a
$k_T$-factorized ansatz involving UGDs. The 
$x$ dependence of those UGDs does not come directly from the 
three-dimensional structure
of the nuclei but from phenomenological fits. While the result in
Ref. \cite{Hirano:2004en} might encode some quantum corrections that are not
included in our work, it also uses a set of assumptions which go beyond
the ones used here.

\begin{table}[t]
\centering
\caption{
The list of parameters for the Woods-Saxon parametrization of 
$d\varepsilon/dy\,[\mathrm{GeV/fm^3}]$ given in Eq.~(\ref{ws}).
Within the approximation of Eq.~(\ref{y-to-eta}), this table can be
seen as a list of parameters for $\varepsilon(\etas)$ as well. Although there is
a large uncertainty in the parameter $\varepsilon_0$, the ratio of 
$\varepsilon_0$ for RHIC to LHC is in compliance with the
ratio that can be calculated for the Stefan-Boltzmann law given the
typical RHIC and LHC temperatures.
  While varying one of the parameters $\as$, $\lambda$ and $\langle \mT \rangle$, we keep the other two fixed to $\as=0.4$, $\lambda=1.8\fm$ and $\langle \mT \rangle=0.7\gev$ for RHIC, and $\as=0.4$, $\lambda=1.8\fm$ and $\langle \mT \rangle=1.3\gev$ for LHC.
}
\begin{ruledtabular}
%%% \scriptsize
\begin{tabular}{ccccccccc}
&&  & RHIC & &  & LHC &   \\
\hline
&  & $\varepsilon_0$ & $\etaflat$ & $a$  & $\varepsilon_0$ & $\etaflat$ & $a$  \\
\hline
$\as$ & 0.2 & 5.8 & 4.6 & 0.7 &  & &  \\
          & 0.3 & 19  & 4.6 & 0.7 & 477 & 6.7 & 0.7 \\
	 & 0.4 & 43 & 4.6 & 0.7 & 1091 & 6.7 & 0.7 \\
         & 0.5 & 77 & 4.6 & 0.7 & 2024 & 6.7 & 0.7 \\
\\
$\lambda$ & 1.4 & 34 &   4.7 & 0.6 & 892 & 6.8 & 0.7\\
                  & 1.6 &  39 & 4.7 & 0.6 & 995 & 6.8 & 0.7\\
                  & 1.8 & 43  & 4.6 & 0.7 & 1091 & 6.7 &0.7 \\
                  & 2.0 & 46  & 4.6 & 0.7 & 1180 & 6.6 & 0.7\\
                  & 2.2 & 50 & 4.5   & 0.7 & 1264 & 6.6 & 0.7 \\
\\
$\langle \mT \rangle$ & 0.5 & 84 & 4.9 & 0.7 & & & \\
           & 0.7 & 43 & 4.6 & 0.7 & & & \\
           & 0.9 & 26 & 4.4 & 0.7 & 2276 & 7.1 & 0.7 \\
           & 1.1 & 17 & 4.2 & 0.7 & 1524 & 6.9 & 0.7 \\
           & 1.3 & 12 & 4 & 0.7 & 1091 & 6.7 & 0.7 \\
           & 1.5 &  & &  &  819 & 6.6 & 0.7 \\
           & 1.7 &  & &  & 638 & 6.4 & 0.7  \\
\end{tabular}
\end{ruledtabular}
\label{table-woods-saxon}
\end{table}

%%%%
%%%%
%%%%
\section{Summary and Outlook}

We have estimated the initial energy density per momentum 
space rapidity $d\varepsilon/dy$ as a function of $y$ at very small 
${\tau\approx 0}$ in the three-dimensional, color neutral McLerran-Venugopalan
model. We have further made an assumption about the correlation 
between the momentum
space rapidity and space-time rapidity, which ultimately allowed us to determine the
energy density as a function of $\etas$. The initial energy density is an important
input, poorly constrained thus far, for current hydrodynamic simulations 
of heavy-ion collisions.
While hydrodynamics sets in at times somewhat later than $\tau \approx 0$, we argue that our results
can nevertheless be
useful as constraints.
The normalization of the energy density depends strongly on some of the 
parameters, but we 
found that for reasonable values of these parameters
the normalizations 
had the correct order of magnitude. Our main result
is the shape of the energy density as a function of rapidity. We provided
a parametrization of our result in terms of a Woods-Saxon function.
Our result is complementary to others found in the literature.

Approaches that first
calculate the number density of gluons have to make assumptions about
how to relate the number of gluons to the energy density as well as the time
at which this matching should be done. We have followed an approach that calculates the
energy density $T^{00}$ directly at a time $\sim R/\gamma$.

It should be straightforward to work the rapidity profile we found here into
the (3+1)$D$ hydrodynamics codes, also including fluctuations
of the energy density in the transverse plane. It would also be interesting
to consider fluctuations in the longitudinal direction, which are beyond the
scope of this work.

%%%%
%%%%
%%%%
\section*{Acknowledgments}
S.O. thanks Joseph Kapusta and Clint Young for discussions. 
S.O. is supported by the U.S. DOE Grant No. DE-FG02-87ER40328 %Minnesota
and 
also partially supported by U.S. DOE Grant No. DE-FG02-00ER41132. %INT
R.J.F. acknowledges support by the U.S. National Science Foundation through 
CAREER Grant PHY-0847538, and by the JET Collaboration and U.S. DOE Grant
DE-FG02-10ER41682.

\bibliographystyle{apsrev4-1}
\bibliography{3dMVn-rapidity}

\end{document}